\documentclass[epj,final]{svjour}
\usepackage{graphics,epsfig,rotating}
\hugehead

\newcommand{\be}{\begin{equation}}
\newcommand{\ee}{\end{equation}}
\newcommand{\bea}{\begin{eqnarray}}
\newcommand{\eea}{\end{eqnarray}}
\begin{document}

\title{On the Space-Time Difference of Proton and Composite Particle Emission 
in Central Heavy-Ion Reactions at 400 $A\cdot$MeV}
\subtitle{\rm FOPI Collaboration}
\author{\rm
R.~Kotte\inst{5} \and H.~W.~Barz\inst{5} 
\and W.~Neubert\inst{5} \and C.~Plettner\inst{5} \and D.~Wohlfarth\inst{5} 
\and J.~P.~Alard\inst{3} \and A.~Andronic\inst{1} 
\and R.~Averbeck\inst{4} \and Z.~Basrak\inst{12} \and 
N.~Bastid\inst{3} \and N.~Bendarag\inst{3} \and
G.~Berek\inst{2} \and R.~\v{C}aplar\inst{12} \and N.~Cindro\inst{12}  
\and P.~Crochet\inst{3} \and A.~Devismes\inst{4} \and 
P.~Dupieux\inst{3} \and M.~D\v{z}elalija\inst{12} \and M.~Eskef\inst{6} 
\and Z.~Fodor\inst{2} \and A.~Gobbi\inst{4} \and Y.~Grishkin\inst{7}
\and N.~Herrmann\inst{6} \and 
K.~D.~Hildenbrand\inst{4} \and B.~Hong\inst{9} 
\and J.~Kecskemeti\inst{2} \and Y.J.~Kim\inst{9} \and M.~Kirejczyk\inst{11}, 
M.~Korolija\inst{12} \and M.~Kowalczyk\inst{11} 
\and T.~Kress\inst{4} \and R.~Kutsche\inst{4} \and A.~Lebedev\inst{7} \and 
K.S.~Lee\inst{9} \and Y.~Leifels\inst{6} \and 
V.~Manko\inst{8} \and H.~Merlitz\inst{6} \and D.~Moisa\inst{1}
A.~Nianine\inst{8} \and
D.~Pelte\inst{6} \and M.~Petrovici\inst{1} \and  
F.~Rami\inst{10} \and W.~Reisdorf\inst{4} \and 
B.~de Schauenburg\inst{10} \and D.~Sch\"ull\inst{4} \and 
Z.~Seres\inst{2} \and B.~Sikora\inst{11} 
\and K.S.~Sim\inst{9} \and V.~Simion\inst{1} \and 
K.~Siwek-Wilczy\'{n}ska\inst{11} \and 
A.~Somov\inst{7} \and G.~Stoicea\inst{1} \and 
M.~A.~Vasiliev\inst{8} \and P.~Wagner\inst{10} \and K.~Wi\'{s}niewski\inst{11} 
\and J.T.~Yang\inst{9} \and Y.~Yushmanov\inst{8} \and A.~Zhilin\inst{7} 
}
\institute{\renewcommand{\thefootnote}{{\rm\alph{footnote}}}
\setcounter{footnote}{0}%
Institute for Nuclear Physics and Engineering, Bucharest, Romania 
\and
Central Research Institute for Physics, Budapest, Hungary 
\and
Laboratoire de Physique Corpusculaire, Universit\'e Blaise Pascal, 
Clermont-Ferrand, France
\and
Gesellschaft f\"ur Schwerionenforschung, Darmstadt, Germany 
\and
Forschungszentrum Rossendorf, PF 510119, 01314 Dresden, Germany,  
\email{kotte@fz-rossendorf.de}
\and
Physikalisches Institut der Universit\"at Heidelberg, Heidelberg, Germany 
\and
Institute for Experimental and Theoretical Physics, Moscow, Russia 
\and
Kurchatov Institute for Atomic Energy, Moscow, Russia 
\and
Korea University, Seoul, Korea
\and
Institut de Recherches Subatomiques, IN2P3-CNRS/ULP, 
Strasbourg, France 
\and
Institute of Experimental Physics, Warsaw University, Warsaw, Poland 
\and
Rudjer Bo\u skovi\'c Institute Zagreb, Zagreb, Croatia 
}
\date{Received: \today}
\titlerunning{Space-Time Difference of Proton and Composite Particle Emission} 
\authorrunning{R.~Kotte et al.}
\abstract{
Small-angle correlations of pairs of nonidentical light charged particles 
produced in central collisions of heavy ions in the $A=100$ mass region  
at a beam energy of 400 $A\cdot$MeV are 
investigated with the FOPI detector system at GSI Darmstadt. 
The difference of longitudinal correlation functions with the 
relative velocity parallel and anti-parallel to the 
center-of-mass velocity of the pair in the central source frame is studied. 
This method allows extracting the apparent 
space-time differences of the emission of 
the charged particles. 
Comparing the correlations with 
results of a final-state interaction model delivers 
quantitative estimates of these asymmetries. 
Time delays as short as 1~fm/c or - alternatively - 
source radius differences of a few tenth fm are resolved. 
The strong collective expansion of the participant zone 
introduces not only an apparent reduction of the source radius 
but also a modification of the emission times. 
After correcting for both effects a complete sequence of the  
space-time emission 
of p, d, t, $^3$He, $\alpha$ particles is presented for the first time.}


\maketitle

\section{Introduction} 
\label{intro}
Correlations of two nonidentical particles 
at small relative velocities are 
- due to final state interactions - sensitive to the 
space-time structure at freeze-out. 
Usually, the correlation is used to extract the size of the source and the 
time duration of the emission 
\cite{Podgoredsky73,Kopylov74,Koonin77,Pratt84,Pratt86,Pratt87,Boal90}. 
However, it contains also information on the 
emission time differences of the two particles. 
Gelderloos and Alexander proposed to construct velocity 
difference spectra at small relative angles \cite{Gelderloos94}. 
Comparing these spectra with results of trajectory model 
calculations they were able to infer the emission order and the time 
intervals between the emission of the two particles \cite{Gelderloos95}.
If the faster particle of a pair approaches and passes the other one 
from behind the pair experiences a stronger final-state interaction 
than in the case that the faster particle has
started in front of the slower one.
Thus, the ratio of the correlation functions with the 
relative velocity parallel and anti-parallel
to the pair velocity is related to the formation sequence.

For heavy-ion collisions of a few hundred MeV per nucleon 
mainly nucleons and light composite particles with charge number $Z \le 2$
are emitted, and one expects relative short time differences. For these
light ejectiles classical trajectory calculations are not
accurate as quantum effects dominate at momenta larger than
$\hbar/a_0$ where $a_0$ is the Bohr radius. 
Very recently, Lednick\'{y} et al. \cite{Lednicky96} have proposed 
to use the above sensitivity of the correlation function 
to directly measure the 
space-time differences in the emission of particles of different types. 
This is possible since the wave functions for nonidentical particles 
are asymmetric with respect to the forward and backward direction  
of the relative momentum. 
First theoretical \cite{Voloshin97} and experimental work 
\cite{Miskowiec98} concentrated on the determination of the asymmetry 
of particle production of proton-pion pairs. 
In the present work the proposed method is 
applied to measure the pair-wise space-time differences of different 
light charged particles. Applying directional cuts
on all relative-velocity correlation functions 
of nonidentical particles allows one to determine - even with a 
certain redundancy - the whole sequence of 
space-time emission points of p, d, t, $^3$He, and $\alpha$ particles.

The two-particle correlations are sensitive to the spatial separation of 
the two particles at the time when the second of them freezes out. 
This separation is the sum of the spatial separation between the two 
freeze-out points, and the time separation, multiplied by the velocity 
of the first particle. A correlation measurement does not 
allow to disentangle these two components.
Therefore, in the following investigation we will discuss two cases, 
first that all particles are emitted from the same region and second that 
they emerge at the same time instant.

\section{The experiment}  \label{experiment}
\subsection{Detector setup}  \label{setup}
The experiment has been performed at the heavy-ion synchrotron SIS at GSI
Darmstadt. Targets of 1~\% interaction thickness of
$^{96}$Ru and $^{96}$Zr have been irradiated by  
$^{96}$Ru and $^{96}$Zr ions of 400$A\cdot$MeV beam energy. 
In order to get sufficient statistics, 
the data of all target-projectile combinations have been used. 
The original aim of the utilization  
of target and projectile nuclei with equal mass but different 
isospin content was to answer the question 
whether the colliding nuclear system attains a full thermo-chemical 
equilibrium during the collision process. The first experimental results  
reveal substantial transparency effects in phase space regions 
already slightly apart of midrapidity \cite{Leifels98}. 

The present analysis uses a subsample of the data, 
taken with the outer Plastic Wall/Helitron 
combination of the FOPI detector system \cite{NIM}. 
The Plastic Wall delivers - via energy loss vs. time-of-flight (TOF) 
measurement - the nuclear charge $Z$ and the velocity ${\bf \beta}$ 
of the particles. The Helitron gives the curvature (which is a measure 
of the momentum over charge $(p/Z)$)   
of the particle track in the field of a large superconducting solenoid.
Since the momentum resolution of the Helitron is rather moderate, 
this detector component serves for particle identification only. The  
mass $m$ is determined 
via  $m c=(p/Z)_{Hel}/(\beta \gamma/Z)_{PlaWa}$, where 
$\gamma=(1-\beta^2)^{-1/2}$. 
The Plastic Wall and the Helitron have full overlap only for 
polar angles between 8.5 degrees and 26.5 degrees. 
The corresponding flight paths amount to 450~cm and 380~cm, 
respectively. 
Monte-Carlo simulations have been performed
in order to study the influence of the finite detector 
granularity and of the TOF and position resolutions  
on the velocity and finally on the proton momentum. 
The resolution of both quantities is governed by the TOF resolution, which 
is $\sigma_{TOF} = 80$~(120)~ps for short (long) scintillator 
strips located at small (large) polar angles \cite{NIM}. 
The detector granularity delivers a negligible contribution 
to the velocity resolution \cite{Kotte97}. 
Thus, between midrapidity ($y_{cm}=0.447$, $\beta_{cm}=0.419$)  
and projectile rapidity ($y_{proj}=0.894$, $\beta_{proj}=0.713$)
the velocity can be determined with a precision 
of $\sigma_{\beta}/\beta \simeq 0.4 \% - 0.8 \%$. 
Finally, from the velocity 
other kinetic quantities like the velocity ${\bf v}^{cm}$ and
the particle momentum ${\bf p}^{cm}$ 
after transformation into the c.m. system of the 
colliding nuclei are deduced. 

\subsection{Event classification} \label{centrality}
\begin{figure}[t]
\vspace*{-17bp}
\centering
\mbox{
\epsfxsize=1.1\linewidth
\epsffile{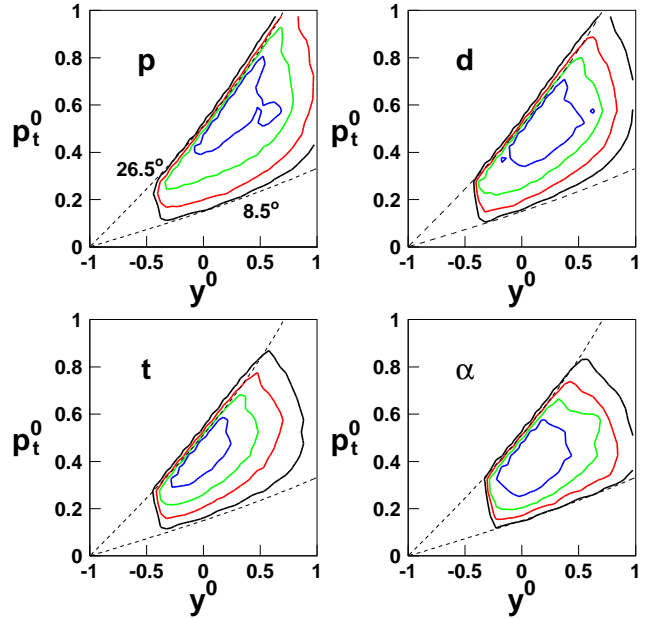}
     }
\caption{
Two-dimensional distribution of yields $d^2\sigma/dp_t dy$ 
of p, d, t, $\alpha$ particles 
in the $p^0_t-y^0$ plane for central reactions selected by a 
8~\% cut on large charged particle multiplicities. 
Target and projectile rapidities are given by $y^0=$~-1 and +1, 
respectively. 
The full lines are levels of constant yield of 20, 40, 60, and 80~\% 
of the maximum value. Dashed lines represent 
the polar angle limits at 8.5 and 26.5 degrees. 
\label{phase_space}
        }
\end{figure}
About $4\cdot 10^6$ central collisions are selected 
by demanding large charged-particle multiplicities to be measured 
in the outer Plastic Wall. The corresponding 
integrated cross section comprises about 8~\% of the total cross section.  
For this centrality class one would expect - within a geometrical picture - 
an average impact parameter of about 2~fm. 
Simulations which we have performed with the IQMD model \cite{Bass} predict
an average impact parameter of about 2.5~fm. 
For different particles with mass number $A_{clus}$   
Fig.\,\ref{phase_space} shows the phase space coverage 
of the detector components, outer Plastic Wall/Helitron, 
in the transverse momentum vs. rapidity plane for events 
selected by this centrality condition. 
Here, $p_t^0=(p_t/A_{clus})/(p_{proj}/A_{proj})_{cm}=
(\beta_t \gamma)/(\beta \gamma)^{proj}_{cm}$ and 
$y^0=(y/y_{proj})_{cm}=(y/y_{cm}-1)$ are the 
normalized transverse momentum and rapidity, respectively. Both observables 
are related to the corresponding projectile quantities in the c.m. frame 
of the colliding nuclei (with $(\beta \gamma)^{proj}_{cm}=0.462$ for
400 $A\cdot$MeV beam energy). It is obvious that for central 
collisions the Plastic Wall preferentially  
measures midrapidity particles with small velocities in the c.m. system 
($\langle v^{cm} \rangle \simeq 0.25~\mbox{c}- 0.30~\mbox{c}$). 

In previous investigations of central Au+Au collisions between 100 and 
400 $A\cdot$MeV beam energy it was found that the  
correlation function of pairs of intermediate mass fragments (IMF) 
is strongly affected by the collective directed sideward flow of nuclear
matter \cite{Kaempfer93,Kotte95}. This directed 
sideflow causes an enhancement of correlations at small relative
momenta. The enhancement results from mixing of differently
azimuthally oriented events; it vanishes if the events are rotated into a 
unique reaction plane, which is determined by the standard   
transverse momentum analysis \cite{Odyniec}. 
This procedure is allowed because the geometrical acceptance and the 
detector efficiency are azimuthally symmetric. 
Thus, the technique of event rotation is applied also 
to the present data in order to prevent that such artificial correlations  
are introduced into the reference momentum distribution of the 
correlation function (cf. Sect.\,\ref{corr_fun}). 

\subsection{Correlation function}\label{corr_fun}
Let $Y_{12}({\bf p}_1, {\bf p}_2)$ be the coincidence yield of pairs 
of particles having momenta ${\bf p}_1$ and ${\bf p}_2$. 
Then the two-particle correlation function is defined as 
\begin{equation}
1 + \mbox{R}({\bf p}_1, {\bf p}_2) = {\cal N} \,
\frac{\sum _{events,pairs} Y_{12}({\bf p}_1, {\bf p}_2)}
{\sum_{events,pairs} Y_{12,mix}({\bf p}_1, {\bf p}_2)}.
\end{equation}
The sum runs over all events fulfilling the above mentioned global 
selection criterion and over all pairs satisfying certain conditions given 
below. 
Event mixing, denoted by the subscript ''mix'',
means to take particle 1 and particle 2 from different events.
${\cal N}$ is a normalization factor fixed by the requirement to have the
same number of true and mixed pairs. 
The statistical errors of all 
the correlation functions presented below   
are governed by those of the coincidence yield, since the mixed yield is
generated with two orders of magnitude higher statistics.
The correlation function (1) 
is then projected onto the relative momentum ${\bf q}$, 
\begin{equation}
{\bf q}=\mu {\bf v}_{12} =\mu ({\bf v}^{cm}_1 - {\bf v}^{cm}_2 ).
\label{defq}
\end{equation}
Here, ${\bf v}^{cm}_{i}$ are the particle velocities calculated 
in the c.m. system of the colliding nuclei and 
$\mu=(m_1 m_2)/(m_1+ m_2)$ 
is the reduced mass of the pair.
Besides the above described global event characteristics
we use gate conditions on the angle $\zeta$ between ${\bf q}$ 
and the c.m. sum momentum of the particle pair    
${\bf P}^{cm}_{12} ={\bf p}^{cm}_1  + {\bf p}^{cm}_2$ 
and on the pair velocity 
$V=\vert {\bf P}^{cm}_{12}\vert/(m_1+m_2)$. 
In order to exploit the full available statistics, two complement types of 
longitudinal correlation functions are generated. 
The forward and backward correlation functions 
are defined by cuts on the angle $\zeta$, 
$\cos{\zeta}>0$ and $\cos{\zeta}<0$, respectively. This choice selects pairs 
with the longitudinal velocity component $v_L$ 
(projection onto the pair velocity) of particle~1 being greater or smaller 
than the corresponding value of particle~2:
\begin{equation}
\mbox{R}^+(q)=1+\mbox{R}(q,\cos{\zeta>0})=1+\mbox{R}(q,v_{L,1}>v_{L,2}) 
\end{equation}
\begin{equation}
\mbox{R}^-(q)=1+\mbox{R}(q,\cos{\zeta<0})=1+\mbox{R}(q,v_{L,1}<v_{L,2}) 
\end{equation}
From the velocity resolution as estimated in 
Sect.\,\ref{experiment} the corresponding $q$ resolution is 
deduced. It is expected to amount to $\delta q=\mu \delta v_{12}$, where 
$\delta v_{12}=\sqrt{2\langle(\delta v^{cm})^2\rangle}=(0.006\pm 0.002)$~c. 

\section{Analysis} \label{analysis}
\subsection{Correlations from final-state interaction} \label{model}
The correlation function of two particles  1 and 2
which move with a pair velocity ${\bf V}$
is \cite{Koonin77,Bauer1} 
\bea
\nonumber
1+R({\bf p_1},{\bf p_2})&=&
\int d t_1 d t_2 d{\bf r}_1 d{\bf r}_2
\rho_1({\bf V},{\bf r}_1,t_1) \rho_2({\bf V},{\bf r}_2,t_2) \times \\
&&\vert\Psi_{\bf q}({\bf r}_1-{\bf r}_2-{\bf V}(t_1-t_2)) \vert^2,
\label{defC}
\eea
where the density $\rho_{1,2}({\bf r},t)$ describes the probability
to find particle 1 or 2, respectively, at time $t$ and spatial
coordinate ${\bf r}$ from which they are emitted (freeze-out configuration). 
The  wave function $ \Psi_{{\bf q}} $
describes the relative motion of the
two particles with momenta ${\bf p}_1$ and ${\bf p}_2$.
Since the interaction with the source is neglected this wave function
depends only on the relative coordinate ${\bf r}= {\bf r}_1-{\bf r}_2$
and on the relative momentum ${\bf q}$ defined in (\ref{defq}).
Here, we assume that the source functions $\rho_{1,2}$
do not depend on the velocity and 
have a Gaussian shape in space and time characterized
by radius parameters $R_{1,2}$ and emission times $\tau_{1,2}$ while
the duration of the emission is given by $\tau_0$:
\begin{equation}
\rho_{1,2}({\bf r},t)=\frac{1}{4 \pi^2 R_{1,2}^3 \tau_0}
\exp[-\frac{r^2}{2 R_{1,2}^2}
-\frac{(t-\tau_{1,2})^2}{2 \tau_0^2}].
\label{defemiss}
\end{equation}
The simple form of Eq.\,(\ref{defemiss}) 
allows us to integrate over the center-of-mass coordinate
$ (m_1 {\bf r}_1 + m_2 {\bf r}_2) / (m_1+m_2)$ and
the time variables leading  to
\begin{equation}
1+R({\bf p_1},{\bf p_2})=
\int d^3r S({\bf r},{\bf V})\vert\Psi_{{\bf q}}({\bf r}) \vert^2 
\label{simpleS}
\end{equation}
with
\bea
\nonumber
S({\bf r},{\bf V}) &=&
\frac{1}{(4\pi)^{3/2}R_0^2 D}
\exp\Big(-\frac{1}{4D^2}\Big[({\bf V}\Delta \tau+{\bf r})^2 \\
&&\phantom{\frac{1}{(4\pi)^{3/2}R_0^2 D}} +
\frac{\tau_0^2}{R_0^2}(r^2 V^2 - ({\bf r \cdot V})^2) 
\Big] \Big),
\label{source}
\eea
where we have introduced the time 
difference
$\Delta \tau_{12}$ and an effective radius $D$ by
\begin{equation}
\Delta \tau_{12}=\tau_1-\tau_2, \quad
D^2=R_0^2+(V \tau_0)^2, \quad
R_0^2=\frac{1}{2}(R_1^2+R_2^2). \quad
\label{dtau}
\end{equation}

For a finite emission time difference $\Delta \tau_{12}$ the 
source function $S$ depends 
on the sign of the scalar product ${\bf V\cdot r}$.
This asymmetry with respect to ${\bf r}$ is transferred to the
correlation function (\ref{simpleS}) if the wave function $\Psi_{\bf q}$
contains terms of an odd power of the scalar product ${\bf q \cdot r}$
which is possible for nonidentical particles.

The wave function $ \Psi_{{\bf q}}$ is generated by partial wave 
expansion technique.
To incorporate the spin degrees of freedom we use either spin-spin coupling
(for p-p, d-p, t-p, $^3$He-p, d-d, t-d, $^3$He-d, and $^3$He-t correlations) 
or l-s coupling 
(for $\alpha$-p, $\alpha$-d, $\alpha$-t, and $\alpha$-$^3$He correlations). 
Thus, the partial waves are classified by angular momentum
and either total spin $s=s_1+s_2$ or $j=l+s_2$.
There is also an option to include partial waves with total angular 
momentum $J=l+s$ to describe dominant resonances.

The corresponding radial Schr\"odinger equations are solved using
the Coulomb potential and Woods-Saxon potentials
\begin{equation}
V(r)=\frac{V_{ws}}{1+\exp[-(r-R_{ws})/a_{ws}]}.
\end{equation}
For each partial wave, 
the parameters $R_{ws}$, $a_{ws}$, V$_{ws}$ are chosen such
that the phase shifts are reproduced. As already 
mentioned in ref. \cite{Boal90} it is important to find
potentials which generate the correct dependence
$d \delta / d q$ since the derivative of the phase shift 
is the relevant quantity in determining 
whether correlations are suppressed or enhanced. 
In most of the cases we make use of the parameters already obtained
in refs.\,\cite{Boal90,Boal86,Jennings86}. 
(Note that for $q$ values approaching a certain resonance positions $q_i$ 
the phase shift has to amount 90 degrees.) 
For partial waves $l>3$ the nuclear part of the potential
is unimportant and has been neglected.
In the case of p-p correlations the wave function has been generated
using the Reid soft-core potential
like in the standard Koonin model \cite{Koonin77}.

The wave function $ \Psi_{{\bf q}}$ in Eq.\,(\ref{defC}) is the projection 
of the many body wave function of the two interacting clusters on their 
relative coordinates. The assumption made above that this wave function 
can be replaced by the partial waves could be violated for small distances 
$\vert {\bf r}_1 -{\bf r}_2 \vert$ where many body effects are most
important. Consequently, the potentials which are obtained by fitting the 
scattering phases might not be the optimum choice to produce the correlation
function. E.g., the d-p correlation functions shown in 
Fig.\,\ref{time_sequence_prot_deut} are too steeply increasing with relative 
momentum $q$. The agreement to the measurement is largely improved using the 
potential depth of 6~MeV instead of the value of -13~MeV given in 
ref.\,\cite{Boal90} for the $s=3/2$, $l=1$ partial wave. However, it is 
important to note that by changing this value the ratio $\mbox{R}^+/\mbox{R}^-$ 
remains essentially unchanged. Therefore, we have decided to use potentials 
which are 
compatible with scattering phases avoiding the ambiguity which could arise by 
searching for new parameters in the very large parameter space. 

\subsection{Effect  of radial flow} 
\label{flow_correct_radius}

From Eqs. (\ref{source}) and (\ref{dtau}) we observe that differences
of the source extensions do not enter in the correlation function.
This fact is a consequence of the isotropic distribution 
of the emission points which is independent of the particle momentum.
However, in central nucleus-nucleus collisions
a considerable amount of the bombarding energy is converted
into flow energy. This collective expansion of nuclear matter
sets in after the compression phase
and can be observed as a decrease of the slope of the kinetic energy spectra.
It causes a strong correlation between particle momenta and emission
points. 
Therefore, we investigate in this section the influence of the
flow on both the source radius and the emission time. 

For this purpose we introduce into the
emission function (\ref{defemiss}) a radial flow velocity.
There is good experience \cite{Reisdorf97} with the "nuclear
Hubble scenario" which suggests a linear
velocity profile. Such an assumption has furthermore the advantage
that the function $S({\bf r},{\bf V})$ can be calculated analytically
when using Gaussian density profiles.
Thus, we write for particles with different mass numbers $A_{1,2}$
\bea
\nonumber
\rho_{1,2}({\bf v},{\bf r},t)&=& N \exp[-\frac{r^2}{2 R_{1,2}^2}
-\frac{(t-\tau_{1,2})^2}{2 \tau_0^2}\\
&&\phantom{N \exp[ }
-\frac{A_{1,2} m_0 ({\bf v}-\eta{\bf r})^2}{2 T}],
\label{newrho}
\eea
where $N$ is a normalization factor which ensures that the
density is normalized to unity. In (\ref{newrho}) the velocities 
of the particles are thermally distributed, 
characterized by the temperature $T$, around a radial flow velocity
defined by the scaling parameter $\eta$.

The "nuclear Hubble constant" $\eta$ is
connected with the radial flow energy per nucleon 
\begin{equation}
\epsilon_{flow}=\frac{m_0 \eta^2}{2}\langle r^2\rangle=
\frac{m_0 \eta^2}{2} 3 R_0^2.
\label{flow_energy}
\end{equation}
The brackets imply averaging over the Gaussian density distribution,
and the quantities $m_0$  and $R_0$
represent the nucleon rest mass and the mean radius, respectively.

Repeating the integration of Eq.\,(\ref{defC}) with the new functions
$\rho_{1,2}$ one arrives at a similar relation for the 
source function $S$ 
as Eq.\,(\ref{source}), however the parameters $R_0$, $D$ and 
$\Delta \tau_{12}$ of Eq. (\ref{dtau}) have to be replaced by quantities 
indicated by an asterisk: 

\be
R^*_0=\sqrt{ \frac{1}{2} (\frac{R_1^2}{1+\epsilon \tilde r_1^2 A_1} +
                          \frac{R_2^2}{1+\epsilon \tilde r_2^2 A_2} )}\,,
\label{radius_flow_correction}
\ee
\be
D^* = \sqrt{(R^*_0)^2 + (V\tau_0)^2} 
\label{d_star}
\ee
and
\be
\Delta \tau_{12}^* = \tau^*_1 - \tau^*_2 = \Delta \tau_{12} + \Delta \tau_{12}^{flow}
\label{delta_tau_star}
\ee
with
\be
\tau^*_{1,2} = \tau_{1,2} - 
\frac{\tilde r_{1,2}^2 A_{1,2}}{1+\epsilon \tilde r_{1,2}^2 A_{1,2}}
\sqrt{\epsilon\frac{m_0}{T}R_0^2}\,,
\label{tau_star}
\ee
and
\be
\Delta \tau_{12}^{flow}=
-\Big(\frac{\tilde r_1^2 A_1}{1+\epsilon \tilde r_1^2 A_1} 
     -\frac{\tilde r_2^2 A_2}{1+\epsilon \tilde r_2^2 A_2} \Big)
\sqrt{\epsilon\frac{m_0}{T}R_0^2}\,.
\label{delta_tau_flow}
\ee
Here, we use the abbreviations $\tilde r_{1,2}=R_{1,2}/R_0$, and 
$\epsilon=\epsilon_{flow}/\epsilon_{therm}$
is the ratio of the radial flow energy $\epsilon_{flow}$
and the energy of the random thermal motion
$\epsilon_{therm}=\frac{3}{2}T$.

From Eq.\,(\ref{radius_flow_correction}) one finds that the apparent 
source radius decreases monotonously with increasing 
energy ratio and particle mass. These observations 
do well compare with recent 
results of the investigation of the sensitivity of the proton-proton 
correlation to collective expansion in central Ni+Ni collisions at 
$1.93\, A\cdot$GeV beam energy \cite{Kotte97}. 

The essential result is that even if the two particles are
produced at the same source region, $R_1=R_2$, and at the same time,
$\tau_1 = \tau_2$, the flow still causes an effective time difference
$\Delta \tau_{12}^* = \Delta \tau_{12}^{flow}$ if the two particle masses
are different. This can be understood as follows: For a given pair
velocity ${\bf V}$ the heavier partner 1 with the smaller thermal velocity
moves approximately with the flow velocity. This means that at
freeze-out the heavier particle is located in the region around
${\bf r}_1={\bf V}/\eta$. Since the position of the lighter partner
with the larger thermal velocity 
is Gaussian distributed around the center of the source, its
mean location ${\bf r}_2$ is behind its heavier partner. Therefore,
to have a strong final state interaction the lighter needs
to start earlier. If they start however at the same time
it seems as if the heavier particle would have started earlier,
$\Delta \tau_{12}^*  < 0$  in agreement with Eqs.\,(\ref{delta_tau_star}) 
and (\ref{tau_star}).

Furthermore, temporal and spatial differences in the source distribution
add up to the effective emission times $\tau^*_i$ in Eq.\,(\ref{tau_star}) 
quite independently. Therefore, measuring the quantity $\Delta \tau_{12}^*$
alone will not allow us to disentangle the two different contributions.
In the following analyses we will therefore differ two cases, namely
first that all particles are emitted from the same source size 
($R_i = R_0$) and 
second that they are emitted at the same time ($\tau_i = 0$) instant.

In our calculations which take the collective expansion into account
we use an energy ratio of $\epsilon=1$ and a temperature of $T=37$~MeV.
These values are compatible with values obtained in
systematic flow studies \cite{Reisdorf97}
of central Au+Au collisions in case of 400 $A\cdot$MeV beam energy.
It should be mentioned that the flow correction $\Delta \tau^{flow}_{12}$
changes only slowly with the energy ratio $\epsilon$ 
because the temperature $T$ in 
Eq.\,(\ref{delta_tau_flow}) is correlated with 
the flow energy $\epsilon_{flow}$ 
due to the requirement of energy conservation. 
Thus, even if for the present Ru+Ru system 
the ratio $\epsilon$ would be smaller 
than for the Au+Au system, one would find that   
the correction changes only marginally. E.g. a drastic 
reduction of $\epsilon$ by a factor of two leads to typical changes   
of $\Delta \tau_{12}^{flow}$ by about 20\% only. 

\section{Results and comparison with model} \label{results}
\subsection{The source extension}\label{source_radius}
\begin{figure}[b]
\centering
\mbox{
\hspace*{-5bp}
\epsfxsize=0.55\linewidth
\epsffile{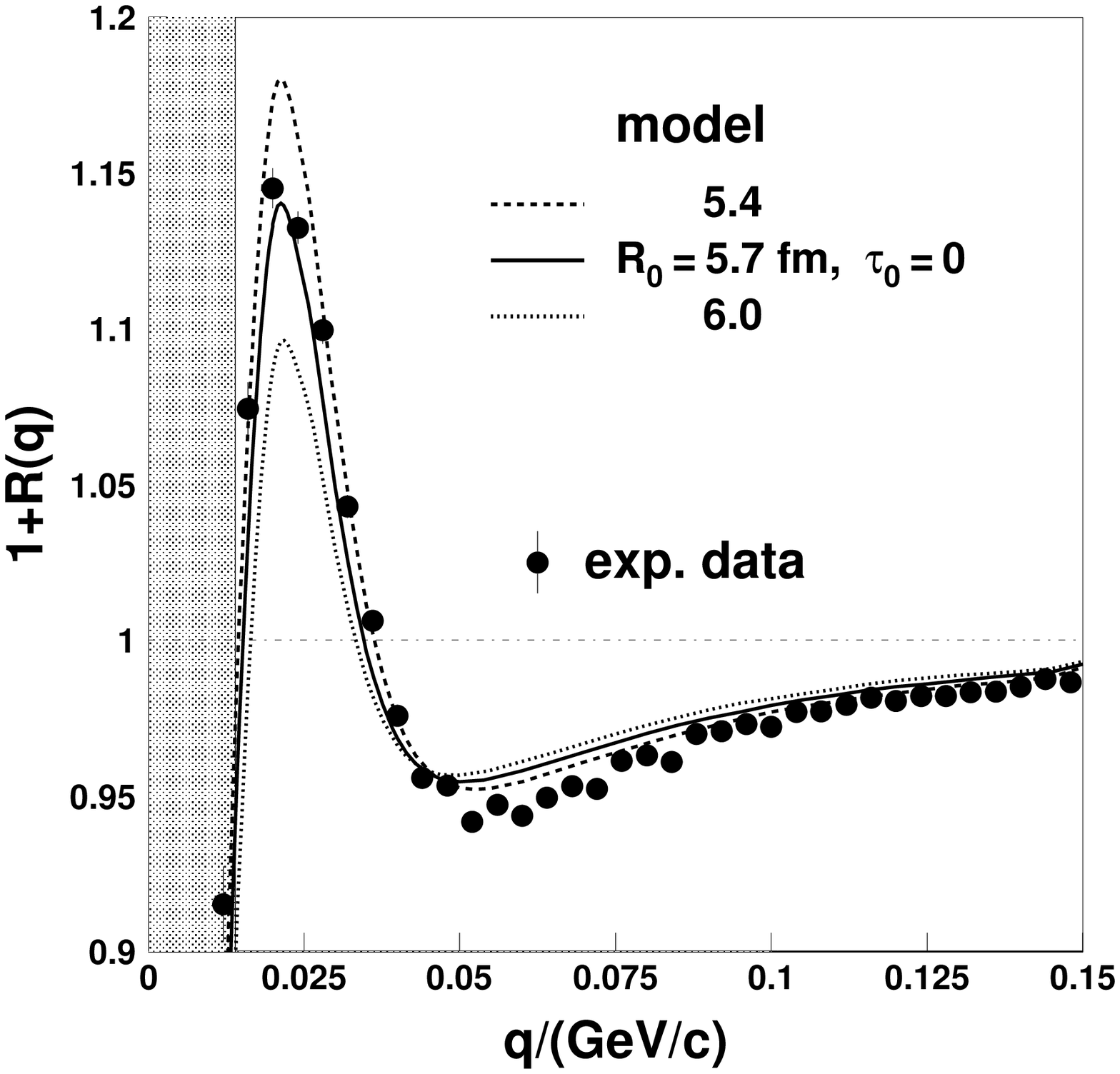} 
\hspace*{-18bp}
\epsfxsize=0.55\linewidth
\epsffile{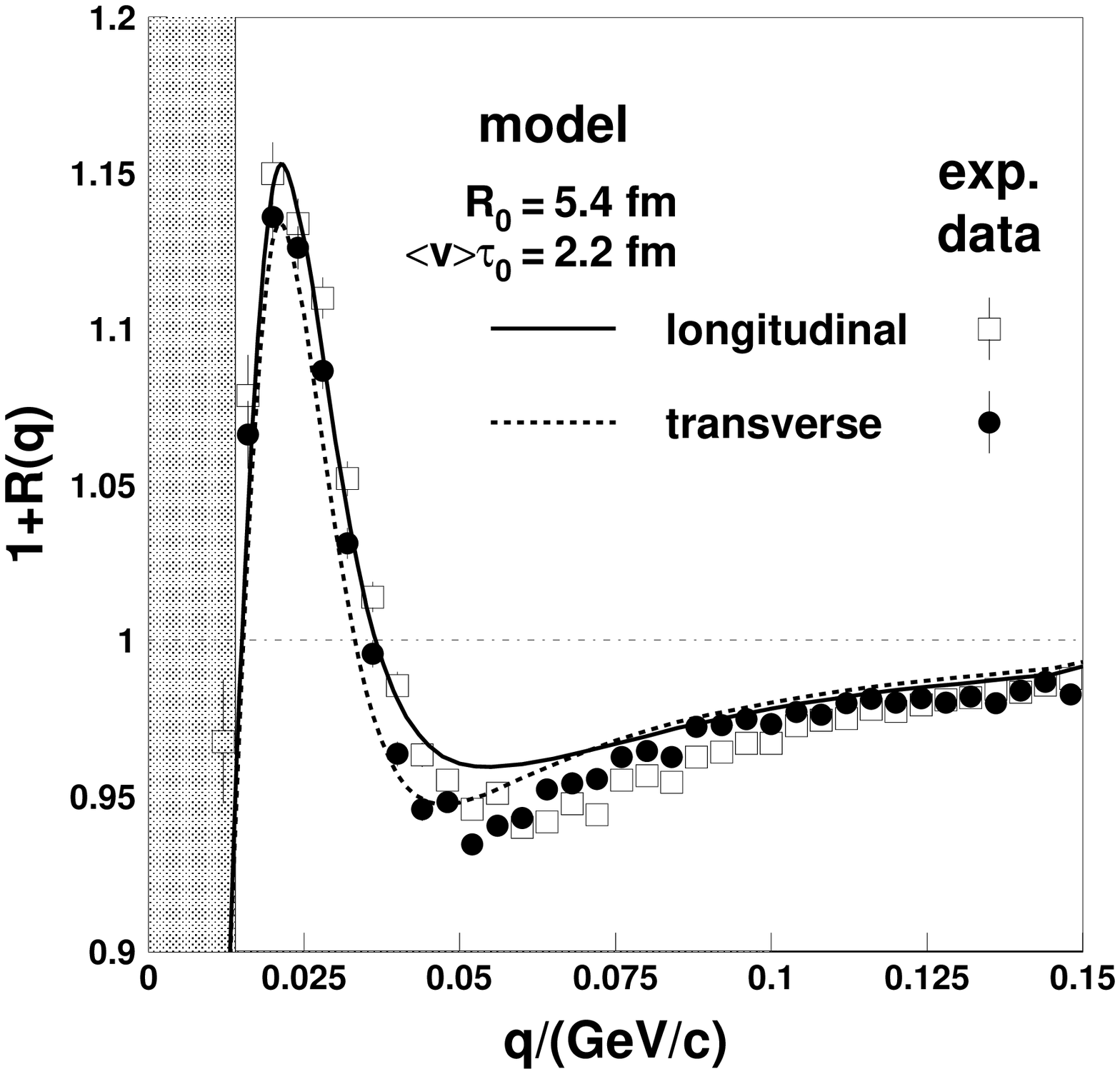}
     }
\caption{Left panel:
The experimental correlation function  
of proton pairs from central trigger 
events (dots). 
The hatched area indicates the unreliable region 
which may be contaminated by doubly counted scattered particles. 
The full line represents the model prediction for  
an emission from a Gaussian source of zero lifetime and 
radius $R_0=5.7$~fm. The dashed and dotted lines are the corresponding 
results for source radii which differ from the optimum one by 
-0.3~fm and +0.3~fm, respectively.
\protect\\
Right panel: 
The transverse (dots) and longitudinal (squares) two-proton correlation 
functions compared to the model results (dashed and dotted lines)
for finite lifetime $\langle V \rangle \tau_0=2.2$~fm and 
Gaussian radius $R_0=5.4$~fm. 
\label{pcor_ru400_hwb}
	}
\end{figure}
Fig.\,\ref{pcor_ru400_hwb} (left panel) shows the angle-integrated 
two-proton correlation function. As in previous fragment-fragment 
correlation analyses \cite{Kaempfer93,Kotte95} an 
enhanced coincidence yield 
at very small relative angles is observed, which is  
due to double counting caused 
mainly by scattering in the scintillator strips. This disturbing yield  
is reduced drastically 
by the requirement to match the particle hits on the Plastic Wall 
with the corresponding tracks in the forward drift chamber Helitron. 
However, mismatches of tracks and scattering processes 
especially at the wire planes of the chamber
can give rise to a small amount of double counting, too. 
The remaining left-over of doubly counted scattered particles
is eliminated by excluding, around a given hit,
positions within a rectangular segment of azimuthal and polar angle 
differences  
$\vert \phi_1 -\phi_2 \vert <4^o$ and $\vert \theta_1 -\theta_2 \vert <2^o$.
In order to keep the influence of the exclusion 
onto the correlation function as small as possible, the same procedure 
is applied to the uncorrelated background. 
However, GEANT simulations \cite{GEANT} have shown that at very small 
relative velocities $v_{12} < 0.03$~c a small  
bias of the correlation function cannot be excluded \cite{Kotte97}. 
Though ratios of correlation functions as our forward/backward relation 
$\mbox{R}^+/\mbox{R}^-$ are expected to be rather robust against such biases, 
the corresponding 
regions in the correlation functions are marked as hatched area and are not
taken into consideration when comparing the experimental data with model 
predictions. 

The experimental p-p correlation function 
is compared to the model for an emission from a source of 
zero lifetime and true Gaussian radius 
$R_0$ which corresponds to an r.m.s. radius 
of $R_{rms}\equiv \sqrt{\langle r^2 \rangle}=\sqrt{3} R_0$.  
The theoretical correlation function is folded with an experimental 
resolution function of Gaussian shape with the  
dispersion $\delta(q)$ as estimated in Sect.\,\ref{corr_fun}. 
The apparent reduction (\ref{radius_flow_correction})
of the source radius due to radial expansion 
effects as described in Sect.\,\ref{flow_correct_radius} (see also 
ref.\,\cite{Kotte97}) is estimated 
to $R_0^*/R_0=1/\sqrt{1+\epsilon}= 1/\sqrt{2}$. 
The best agreement of experimental data and model 
calculations is found for 
$R_0=(5.7 \pm 0.3)$~fm. It is obvious that the appearance of the 
correlation peak at $q\simeq 20$~MeV/c allows the determination of  
the source radius with rather high sensitivity. This $^2$He-resonance  
is the result of the common action of the enhancement due to the  
attractive nucleonic potential and the suppression due to both 
the mutual Coulomb repulsion 
and the antisymmetrization of the wave function. 
The ambiguity of the space and time extents of the proton 
emitting source can be resolved by constructing correlation functions with the 
relative momentum ${\bf q}$ being either perpendicular (transverse 
correlation function, here defined by $\vert \cos \zeta \vert <0.5$)  
or parallel (longitudinal correlation function, 
here defined by $\vert \cos \zeta \vert >0.5$) 
to the pair velocity ${\bf V}$. 
The result is given in the right panel of Fig.\,\ref{pcor_ru400_hwb}. 
We find a small suppression of the transverse correlations with respect
to the longitudinal ones which is consistent with model predictions 
for the emission from a source of finite lifetime 
\cite{Koonin77,Pratt87}. The best simultaneous fit around the
correlation peaks ($\mbox{14 MeV/c}<q<\mbox{42 MeV/c}$) of both the transverse 
and longitudinal correlation functions delivers Gaussian parameters of the radius 
(after flow correction) and the emission duration of $R_0=5.4$~fm and 
$\langle V \rangle \tau_0=2.2$~fm, 
respectively. Obviously, the lifetime effect appears unimportant. 
  
For the determination of the emission time differences 
in the subsequent section 
the source radius is kept fixed to a value of $R_1=R_2=R_0=5.7$~fm.  
For simplification the duration of the emission $\tau_0$ is set to zero.  
A finite value leads only to a rescaling of the effective radius 
in Eq.\,(\ref{d_star}).

\subsection{Emission time differences}\label{time_differences}
Out of the ten different combinations of pairs of nonidentical 
particles which can be constructed from p, d, t, $^3$He, and $\alpha$ 
particles there exist only two correlation functions which do not contain 
contributions of resonance decay products. 
These resonance-free correlation functions  
of deuteron-proton and $^3$He-triton pairs are presented 
in Sect.\,\ref{resonance_free_correlations}. All other 
correlation functions contain resonance contributions of   
particle-unbound ground states or excited states of heavier clusters 
decaying into pairs of light charged particles. 
In case of the appearance of strong and narrow resonances 
in the $q$ region of interest, the corresponding 
theoretical correlation functions are corrected for 
the experimental $q$ resolution similarly as for p-p correlations. 

\subsubsection{Final-state interaction without resonances} 
\label{resonance_free_correlations}
\begin{figure}[t]
\vspace*{-17bp}
\centering
\mbox{
\epsfxsize=1.\linewidth
\epsffile{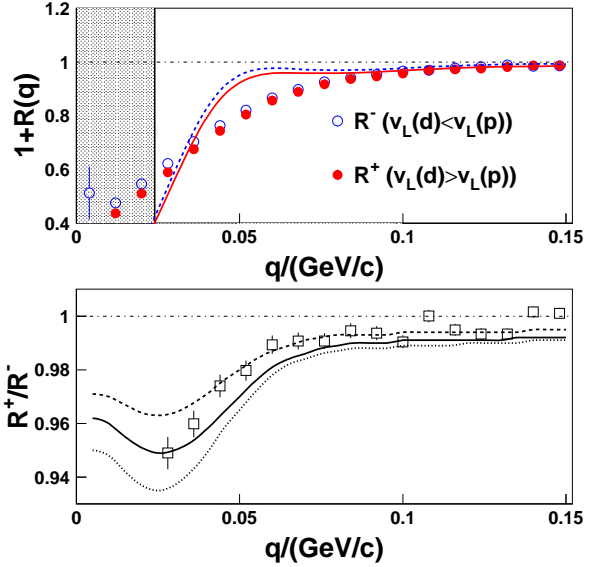} 
     }
\caption{
Upper panel: Forward (full dots) and backward (open dots) 
longitudinal experimental correlation functions of d-p pairs. 
The hatched area indicates the unreliable region 
which may be contaminated by doubly counted scattered particles. 
The full and dashed lines
give the corresponding model predictions with the time delay of 
table\,\protect\ref{pair_wise_time_delays}. 
Lower panel: Ratio of forward/backward experimental correlation 
functions (open squares). The full line represents  
the ratio of the simulated correlation functions. 
The dashed and dotted lines give the model predictions for times 
differences deviating by -0.5~fm/c and +0.5~fm/c from the optimum one,
respectively. 
\label{time_sequence_prot_deut}
	}
\end{figure}
\begin{figure}[b]
\vspace*{-17bp}
\centering
\mbox{
\epsfxsize=1.\linewidth
\epsffile{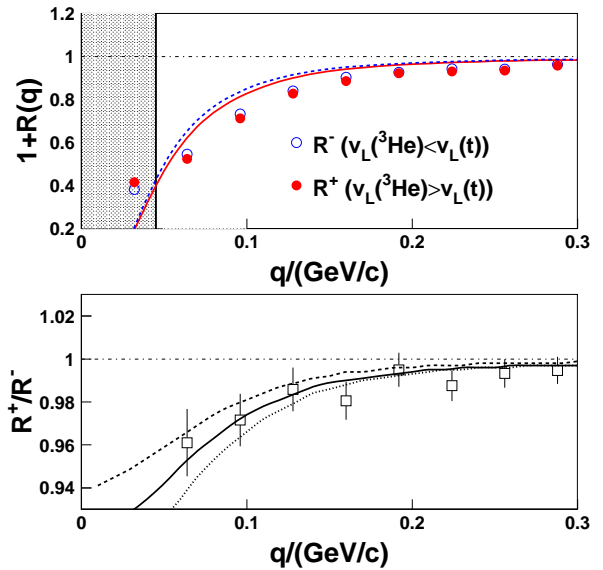} 
     }
\caption{
The same as Fig.\,\protect\ref{time_sequence_prot_deut}, but for 
$^3$He-t correlations. 
\label{time_sequence_3he_trit}
	}
\end{figure}
Fig.\,\ref{time_sequence_prot_deut} gives the results of d-p
correlations. The upper panel shows the forward (full dots) and the 
backward (open dots) correlation functions whereas the lower panel 
represents the ratio of both observables. The resonance-free correlation 
functions show a suppression at low relative momenta due to final-state 
Coulomb and nuclear interactions. Obviously, the suppression gets stronger 
in the case when deuterons are faster than protons 
(full dots in Fig.\,\ref{time_sequence_prot_deut}). 
This indicates that, on average, deuterons are being emitted 
later than protons.
Quantitatively, the comparison with the model predictions yields an 
optimum time delay of $\Delta \tau_{\mbox{d,p}}= \tau_{\mbox{d}} - 
\tau_{\mbox{p}}=6.5$~fm/c (full line in the lower panel). 
(The dashed and dotted lines in the lower panel of 
Figs.\,\ref{time_sequence_prot_deut} 
and \ref{time_sequence_3he_trit} should give an impression 
how the theoretical ratio $\mbox{R}^+/\mbox{R}^-$ alters if the  
time delay is changed by -0.5~fm/c and +0.5~fm/c, respectively.)
However, if one does not take into account the 
time shift due to the radial expansion of the participant zone  
the emission time difference would be determined as an apparent value 
which is much smaller (but still positive). 
Both the true and apparent time delays are summarized in 
table\,\ref{pair_wise_time_delays}. 
The given errors represent the typical band widths of time delay parameters  
around their optimum values which reproduce the 
experimental correlation function ratios $\mbox{R}^+/\mbox{R}^-$ 
reasonably well. 
\begin{table}[b]
\vspace*{10bp}
\caption{The apparent and the true values of the emission time difference 
as determined for all combinations of nonidentical charged-particle 
correlation functions. The source radii are fixed to $R_i=R_0$.
\label{pair_wise_time_delays}
	}
{\setlength{\tabcolsep}{5.5mm}     
\begin{tabular}{rrr}
\hline
light-charged & apparent  & true 
\protect\\
particle & time delay  & time delay
\protect\\
combination & $\Delta \tau_{12}^*$  &  $\Delta \tau_{12} = \tau_1 -\tau_2$
\protect\\
1 - 2 & (fm/c) & (fm/c)
\protect\\ \hline
d - p & $ 1.7 \pm 1$ & $6.5 \pm 1$
\protect\\
t - p & $ -4.2 \pm 3$ & $3 \pm 3$  
\protect\\
$^3$He - p & $4.5 \pm 1$ & $11.7 \pm 1$ 
\protect\\ 
$\alpha$ - p & $-2.6 \pm 3$ & $ 6 \pm 3$
\protect\\ 
t - d & $ -1.0 \pm 1$ & $1.4 \pm 1$ 
\protect\\ 
$^3$He - d  & $3.0 \pm 1$ & $5.4 \pm 1$
\protect\\ 
$\alpha$ - d  & $-2.0 \pm 1$ & $1.8 \pm 1$
\protect\\
$^3$He - t & $1.7 \pm 1$ & $1.7 \pm 1$  
\protect\\ 
$\alpha$ - t & $ -2.9 \pm 1$ & $-1.5 \pm 1$
\protect\\ 
$^3$He - $\alpha$ & $4.4 \pm 1$ & $3.0 \pm 1$
\protect\\ \hline
\end{tabular}
}
\end{table}

Fig.\,\ref{time_sequence_3he_trit} shows the $^3$He-t correlation functions 
and the corresponding forward/backward ratio. 
Since the 
masses of the two species are practically identical, 
this correlation function is the only one where 
- in case of equal source radii - 
the flow correction (\ref{delta_tau_flow}) vanishes 
and, consequently, 
the true and the apparent time delays are identical. 
Obviously, the $^3$He particles are emitted slightly later than the tritons 
(about 1-3~fm/c, cf. table\,\ref{pair_wise_time_delays}). 

\subsubsection{Final-state interaction with resonances} 
\label{resonance_containing_correlations}
\begin{figure}[t]
\vspace*{-17bp}
\centering
\mbox{
\epsfxsize=1.\linewidth
\epsffile{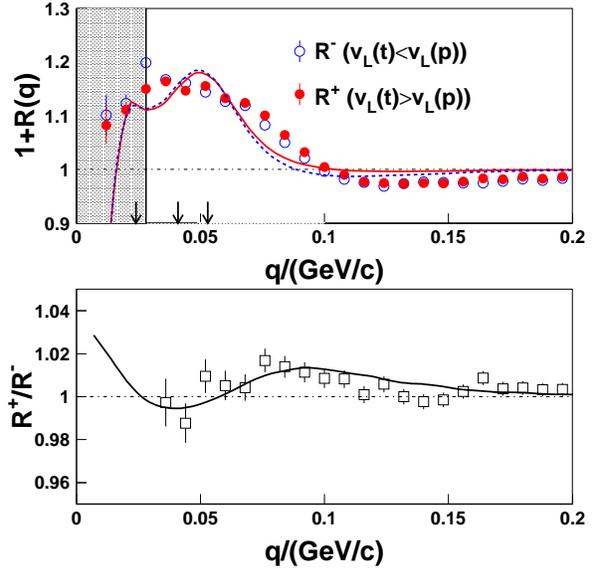} 
     }
\caption{
Upper panel: Forward (full dots) and backward (open dots) 
longitudinal experimental correlation functions of t-p pairs. 
Positions of relevant resonances are marked as arrows. 
The hatched area indicates the unreliable region 
which may be contaminated by doubly counted scattered particles. 
The full and dashed lines 
give the corresponding model predictions with the time delay of 
table\,\protect\ref{pair_wise_time_delays}.   
Lower panel: Ratio of forward/backward experimental correlation 
functions (open squares). The full line represents  
the ratio of the simulated correlations. 
\label{time_sequence_prot_trit}
	}
\end{figure}

\begin{figure}[t]
\vspace*{-17bp}
\centering
\mbox{
\epsfxsize=1.\linewidth
\epsffile{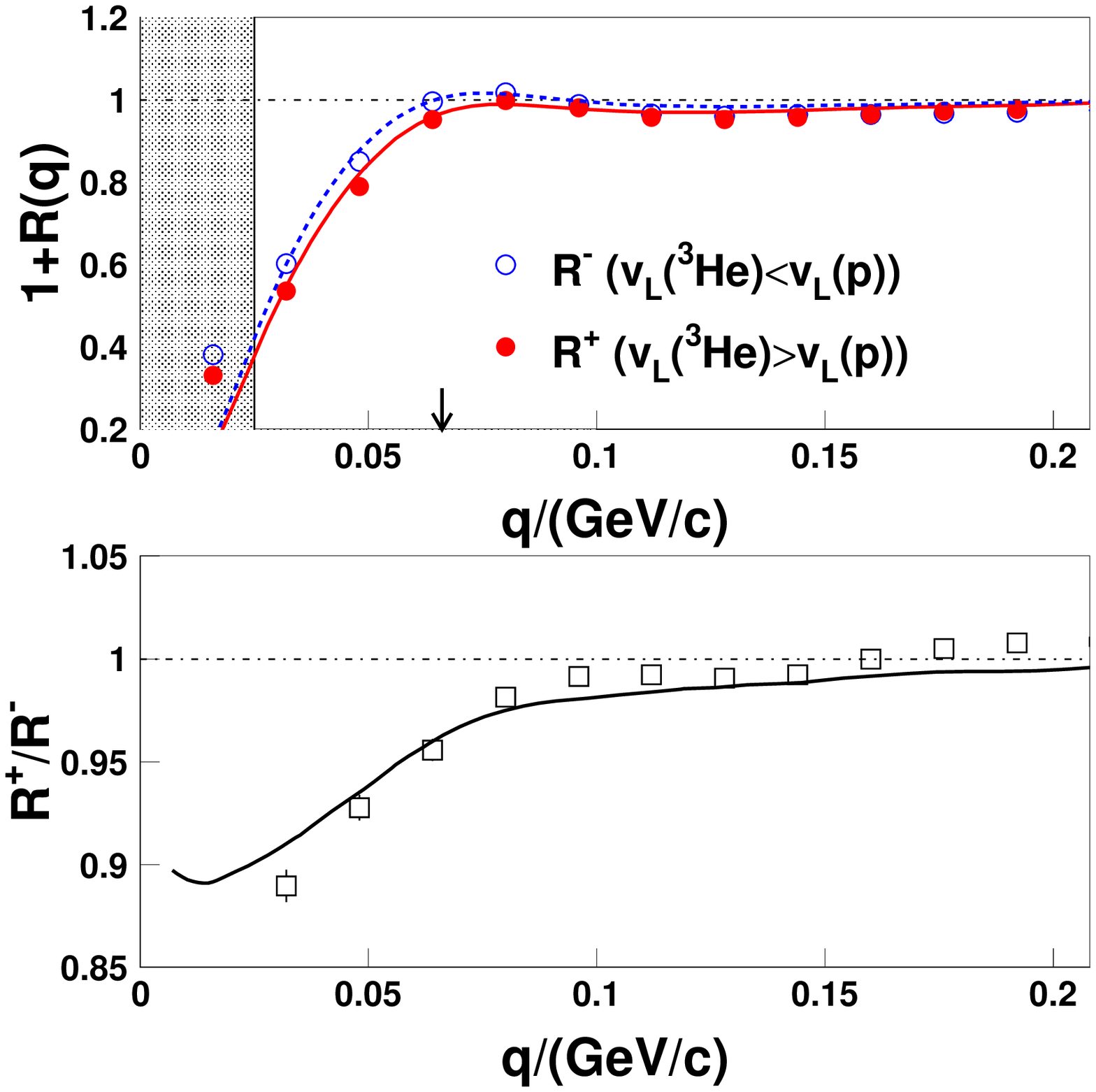} 
     }
\caption{
The same as Fig.\,\protect\ref{time_sequence_prot_trit}, but for 
$^3$He-p correlations. 
\label{time_sequence_3he_prot}
	}
\end{figure}

\begin{figure}[t]
\vspace*{-17bp}
\centering
\mbox{
\epsfxsize=1.\linewidth
\epsffile{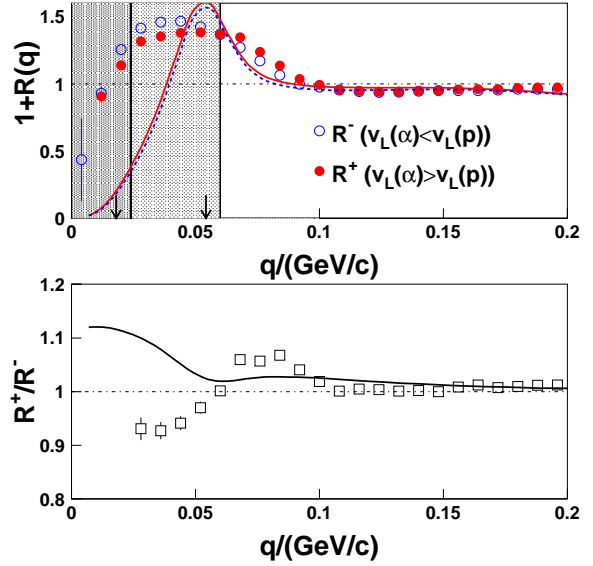} 
     }
\caption{
The same as Fig.\,\protect\ref{time_sequence_prot_trit}, but for 
$\alpha$-p correlations. 
The dark hatched area indicates the unreliable region 
which may be contaminated by doubly counted scattered particles.   
The light hatched area gives the region which 
is supposed to be strongly contaminated by secondary decays of boron isotopes. 
\label{time_sequence_4he_prot}
	}
\end{figure}

\begin{figure}[b]
\vspace*{-17bp}
\centering
\mbox{
\epsfxsize=1.\linewidth
\epsffile{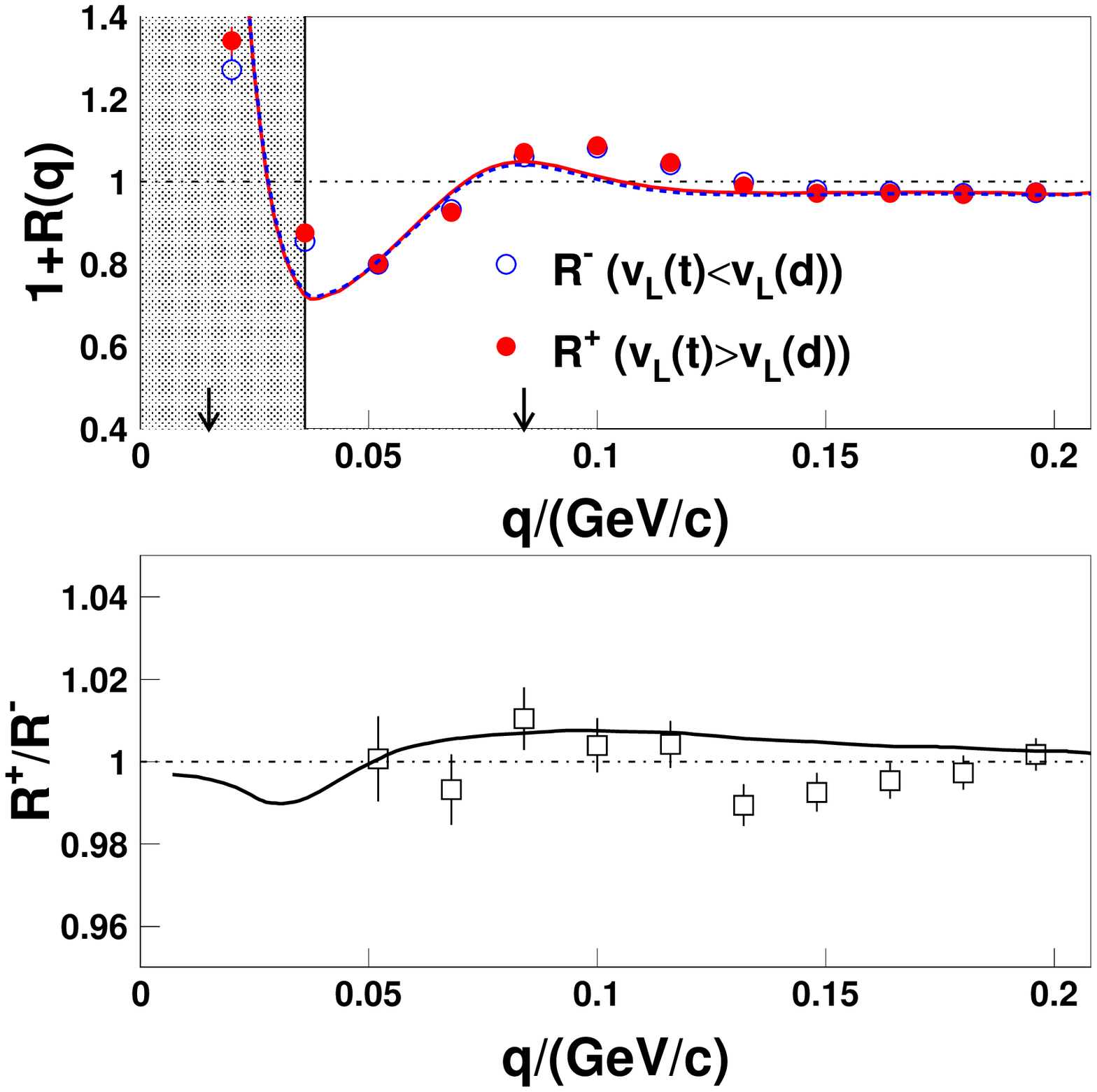} 
     }
\caption{
The same as Fig.\,\protect\ref{time_sequence_prot_trit}, but for 
t-d correlations. 
\label{time_sequence_trit_deut}
	}
\end{figure}

\begin{figure}[t]
\vspace*{-17bp}
\centering
\mbox{
\epsfxsize=1.\linewidth
\epsffile{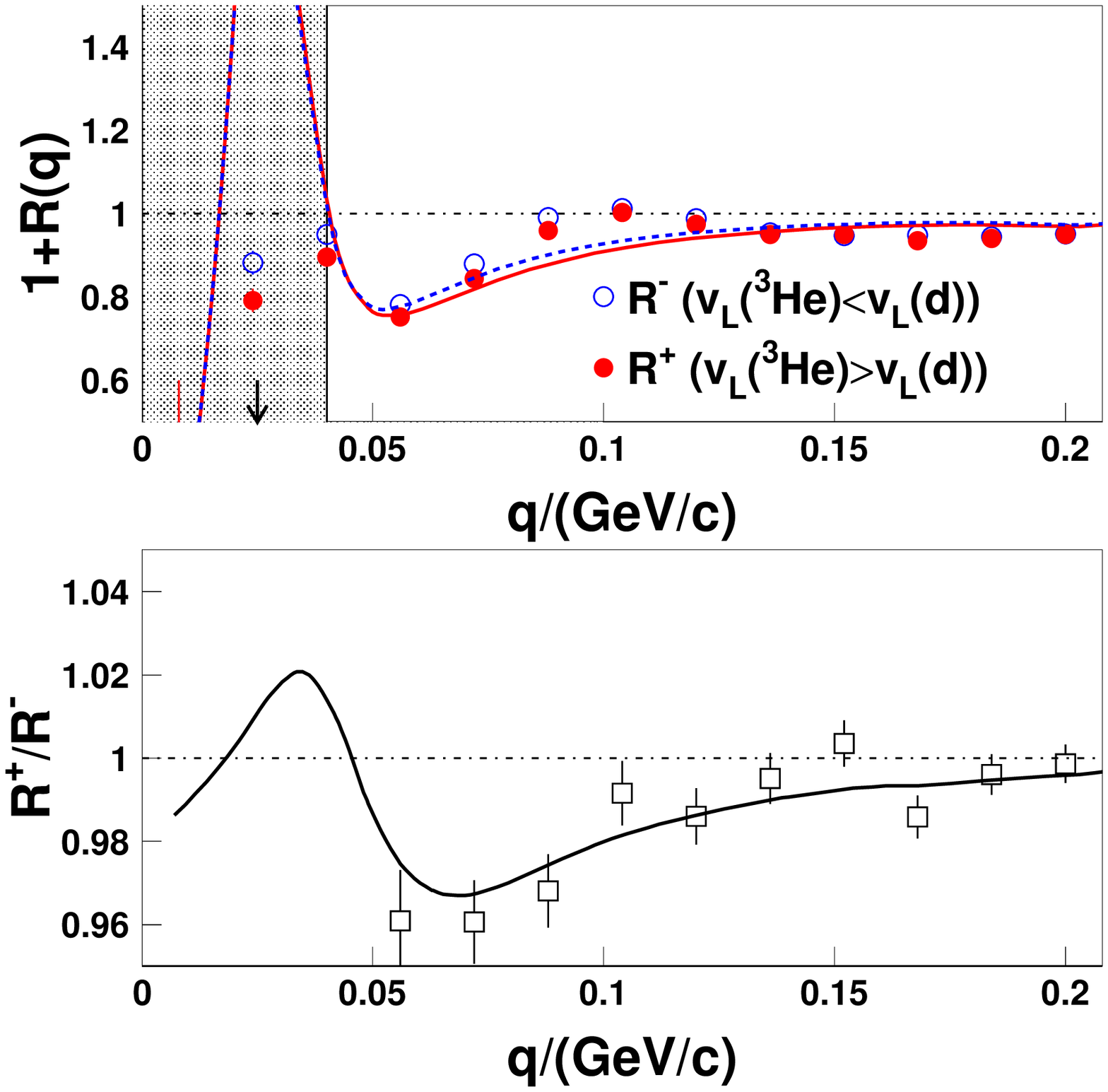} 
     }
\caption{
The same as Fig.\,\protect\ref{time_sequence_prot_trit}, but for 
$^3$He-d correlations. 
\label{time_sequence_3he_deut}
	}
\end{figure}

\begin{figure}[t]
\vspace*{-17bp}
\centering
\mbox{
\epsfxsize=1.\linewidth
\epsffile{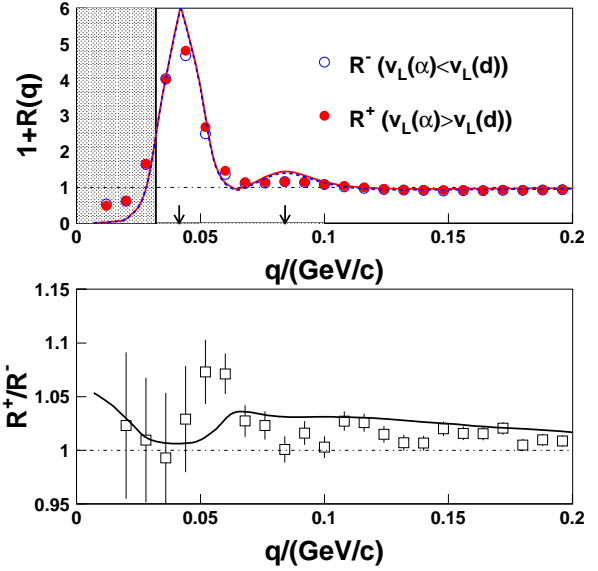} 
     }
\caption{
The same as Fig.\,\protect\ref{time_sequence_prot_trit}, but for 
$\alpha$-d correlations. 
\label{time_sequence_4he_deut}
	}
\end{figure}

\begin{figure}[b]
\vspace*{-17bp}
\centering
\mbox{
\epsfxsize=1.\linewidth
\epsffile{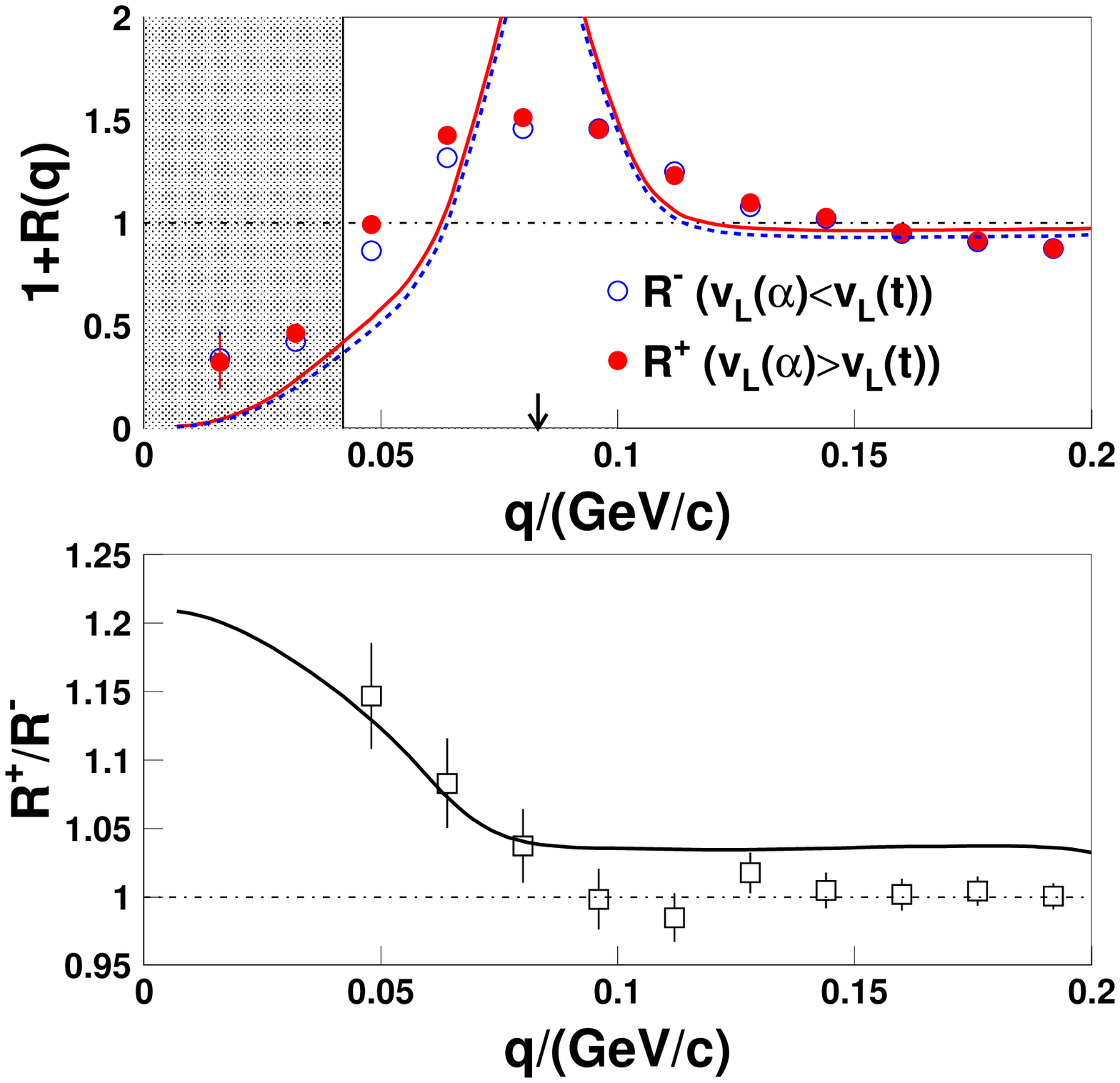} 
     }
\caption{
The same as Fig.\,\protect\ref{time_sequence_prot_trit}, but for 
$\alpha$-t correlations. 
\label{time_sequence_4he_trit}
	}
\end{figure}

\begin{figure}[t]
\vspace*{-17bp}
\centering
\mbox{
\epsfxsize=1.\linewidth
\epsffile{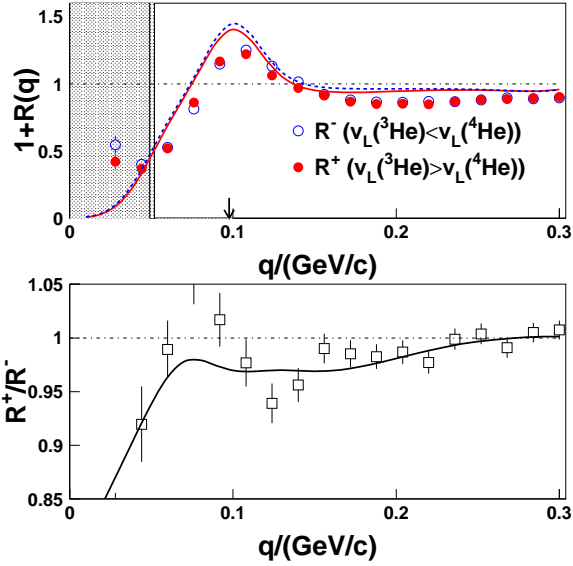} 
     }
\caption{
The same as Fig.\,\protect\ref{time_sequence_prot_trit}, but for 
$^3$He-$\alpha$ correlations. 
\label{time_sequence_3he_4he}
	}
\end{figure}

If a resonance contribution is dominating in the two-particle yield  
the ratio $\mbox{R}^+/\mbox{R}^-$ necessarily is forced 
to unity for relative momenta 
$q$ approaching the resonance value $q_i$. This is due to the fact that 
for pure two-body decays both particles are emitted at the same 
time (and position). Indeed, the experimental data show the expected 
behaviour (see arrow positions in 
Figs.\,\ref{time_sequence_prot_trit}-\ref{time_sequence_3he_4he}).

Fig.\,\ref{time_sequence_prot_trit} shows the forward/backward 
longitudinal correlation functions of t-p pairs  
together with the corresponding ratio. The correlation function exhibits 
a broad peak which contains the contribution of the 1st excited state
of $^4$He ($E^*=20.21$~MeV, $J^{\pi}=0^+$, $\Gamma=0.5$~MeV, 
$\Gamma_p/\Gamma=1$, $q_1=23.6$~MeV/c) 
as well as the 2nd ($E^*=21.01$~MeV, $J^{\pi}=0^-$, $\Gamma=0.84$~MeV,
$\Gamma_p/\Gamma=0.76$, $q_2=41.0$~MeV/c)
and the 3rd one ($E^*=21.84$~MeV, $J^{\pi}=2^-$,
$\Gamma=2.01$~MeV, $\Gamma_p/\Gamma=0.63$, $q_3=53.4$~MeV/c). 
The positions of the relevant resonances are marked by arrows. 
In the present case only the lowest three excited states of $^4$He
with widths $\Gamma<2.1 $~MeV are indicated. Other states 
in the energy region $E^*=23-26$~MeV ($q=70-90$~MeV/c) are much broader   
($\Gamma>5$~MeV, $\Gamma_p/\Gamma\simeq 0.5$). 
Obviously, from the lower part of the figure 
one would conclude a negative time delay 
$\Delta \tau^*_{\mbox{t,p}}$. However, the radial flow correction
(\ref{delta_tau_flow}) overcompensates 
this apparent time difference and 
leads to a positive value (cf. table\,\ref{pair_wise_time_delays}).
Due to the possible contribution of higher lying resonances of $^4$He  
which are not taken into account in the model 
description, a rather large error is appended to the time delay.

Fig.\,\ref{time_sequence_3he_prot} represents the results 
for $^3$He-p correlations. One broad maximum shows up which is 
related to the particle-unbound ground state of $^4$Li 
($J^{\pi}=2^-$, $\Gamma=6$~MeV, $q_0=66.0$~MeV/c). An unquestionably positive 
time delay $\Delta \tau_{^3\mbox{He,p}}$ 
is deduced from the correlation function 
ratio $\mbox{R}^+/\mbox{R}^-$. Here, about 60~\% of the true time difference  
given in table\,\ref{pair_wise_time_delays} arise 
from the radial flow correction (\ref{delta_tau_flow}). 
  
Fig.\,\ref{time_sequence_4he_prot} gives the $\alpha$-p correlations. 
The origin of the broad bump is not fully understood. 
The main contribution results from the decay of the particle-unbound 
ground state of $^5$Li ($J^{\pi}=\frac{3}{2}^-$, 
$\Gamma=1.5$~MeV, $\Gamma_p/\Gamma=1$, $q_0=54.3$~MeV/c). 
Most probably, the bump contains additional contributions  
\cite{Pochodzalla85,ZhiYongHe97,Charity95} 
which cannot be separated experimentally from the $^5$Li resonance. 
One contribution is expected at $q\simeq 16 $~MeV/c. 
It corresponds to the three-body decay of 
$^9\mbox{B(g.s.)}\rightarrow$ p + $^8$Be(g.s.) $\rightarrow$ p + $\alpha$ + $\alpha$, 
where only one of the $\alpha$ particles is detected together with the proton. 
In addition, the four-body decay of 
$^{10}$B$^* \rightarrow$ p + $^9$Be(1.69 MeV) $\rightarrow$
p + n + $^8$Be(g.s.) $\rightarrow$ p + n + $\alpha$ + $\alpha$
can contribute to the broad maximum 
as was deduced at a comparably low beam energy of 40 $A\cdot$MeV
\cite{Charity95}. In the present case such a contribution cannot be ruled out
since a considerable production of 0.3 boron clusters per event was found 
in central Au+Au reactions at 400 $A\cdot$MeV beam energy \cite{Reisdorf97}.

Due to the large lifetime of the $^8$Be ground state ($\Gamma =6.8 $~eV) 
the protons - a priori - are emitted earlier than the $\alpha$ particles. 
Thus, for relative momenta $q\simeq 0 - 60 $~MeV/c 
one necessarily expects stronger final-state interaction if 
$v_L(\alpha)>v_L(\mbox{p})$ and consequently $\mbox{R}^+/\mbox{R}^-<0$.
Indeed, the experimental data follow the predicted trend. However, 
since the model description does not incorporate two-stage decays, 
a reliable time difference can only be extracted for $q$ values well above the 
resonance $q_0$. Keeping in mind that the 
sensitivity of the correlation function to a variation of the model 
parameter $\Delta \tau_{12}$ decreases for increasing $q$, the 
deduced value is affected with a rather large error. 
Thus, the time difference derived from the $\alpha$-p 
correlation function will enter with negligible weight into the procedure 
used to determine the emission-time sequence of the different particle species 
(cf. Sect.\,\ref{discussion}).

Two resonances affect the t-d correlation function 
(Fig.\,\ref{time_sequence_trit_deut}). The first one which is not resolved 
experimentally corresponds to the 16.75~MeV excited state of $^5$He 
($J^{\pi}=\frac{3}{2}^+$, $\Gamma=76$~keV, $q_1=10.8$~MeV/c). The second one 
is due to the state at 19.8~MeV ($J^{\pi}=(\frac{3}{2},\frac{5}{2})^+$, 
$\Gamma=2.5$~MeV, $q_2=83.5$~MeV/c). From the lower part of the figure it is 
obvious that the apparent time delay is close to zero. Indeed, 
the model fits well the data for a true emission time difference 
(cf. table\,\ref{pair_wise_time_delays})
which - to a large extent - is due to the contribution of the radial 
flow correction (\ref{delta_tau_flow}).

The $^3$He-d correlations in 
Fig.\,\ref{time_sequence_3he_deut} appear very similar to the 
t-d correlations. The correlation function 
contains one strong peak due to 
the 16.66~MeV state of $^5$Li ($J^{\pi}=\frac{3}{2}^+$, 
$\Gamma=0.3$~MeV, $q_1=24.8$~MeV/c). 
Other broader states are at 
18.0~MeV ($J^{\pi}=\frac{1}{2}^+$, $\Gamma\simeq 5$~MeV, $q_2=60.3$~MeV/c)
and at 20.0~MeV ($J^{\pi}=(\frac{3}{2},\frac{5}{2})^+$, 
$\Gamma= 5$~MeV, $q_3=90.2$~MeV/c). 
Only the narrow 16.66~MeV state is taken into account in the model.
In contrast to the t-d correlations, 
we find a clearly positive apparent time delay 
$\Delta \tau_{^3\mbox{He,d}}^* \simeq 3$~fm/c  
which increases nearly by a factor of 2 due to the radial flow 
correction. 

The $\alpha$-d correlation function shown in the 
upper panel of Fig.\,\ref{time_sequence_4he_deut} is governed by 
a strong resonance due to the narrow 2.186~MeV state of $^6$Li 
($J^{\pi}=3^+$, $\Gamma=24$~keV, $q_1=42.2$~MeV/c). Another state
is at 4.31~MeV ($J^{\pi}=2^+$, $\Gamma=1.7$~MeV, $q_2=84.2$~MeV/c). 
Obviously, from the lower part of the figure 
one would conclude a negative 
time delay $\Delta \tau^*_{\alpha,\mbox{d}}$. However, 
for the present case the radial flow correction
(\ref{delta_tau_flow})  
overcompensates the apparent time difference  
leading to a positive value (cf. table\,\ref{pair_wise_time_delays}).
As a by-product, the strong $3^+$ resonance of $^6$Li in the 
correlation function can serve for an independent determination of the 
$q$ resolution. A Gaussian fit to the difference 
spectrum of true and normalized mixed yields 
in the region 20~MeV/c~$<q<$60~MeV/c delivers a dispersion of 
$\delta v_{12} =\delta q / \mu = 
7.1~\mbox{(MeV/c)}/\mu=0.0057$~c 
in good agreement with the estimate given in Sect.\,\ref{corr_fun}.

The $\alpha$-t correlation function given in 
Fig.\,\ref{time_sequence_4he_trit} is characterized by a broad maximum 
due to the 4.63~MeV state of $^7$Li ($J^{\pi}=\frac{7}{2}^-$, 
$\Gamma=93$~keV, $q_1=83.4$~MeV/c). Other states are at 
6.68~MeV ($J^{\pi}=\frac{5}{2}^-$, $\Gamma=880$~keV, $q_2=116.4$~MeV/c)
and at 7.46~MeV ($J^{\pi}=\frac{5}{2}^-$,
$\Gamma=89$~keV, $q_3=126.7$~MeV/c). Only the 4.63~MeV state  
is taken into account in the model. 
The deduced time delay $\Delta \tau_{\alpha,\mbox{t}}$ 
is apparently negative; it is reduced by about 50~\% when taking into account 
the radial flow correction (cf. table\,\ref{pair_wise_time_delays}).

The $^3$He-$\alpha$ correlation function 
in Fig.\,\ref{time_sequence_3he_4he} is dominated by the 4.57~MeV 
state of $^7$Be ($J^{\pi}=\frac{7}{2}^-$, $\Gamma=175$~keV, $q_1=98.0$~MeV/c). 
Here, both the deduced apparent and true time delays 
$\Delta \tau_{^3\mbox{He},\alpha}$ are found positive 
(table\,\ref{pair_wise_time_delays}) and differ 
only by about 1.5~fm/c. Similarly to the $\alpha$-t correlations 
the small flow correction is a result of the small relative 
mass difference of the particles. 

\subsection{Discussion}\label{discussion}
Finally, the redundancy of the ten different pair-wise time 
differences obtained above 
allows one to fix rather reliably the  
emission-time sequence of the light charged particles. The optimum sequence
is derived from a least-squares solution of the set of  
linear equations for the time delays given in 
table\,\ref{pair_wise_time_delays}. 
(Note that, the deduced $\chi^2$ per degree
of freedom of 1.1 does not carry quantitative information 
on the quality of the regression since the errors of the individual time delays 
are not real standard deviations.) 
The different contributions are 
weighted by the inverse squares of the given errors.  
If one excludes from the fit a few of the time differences 
(e.g. those which are affected with large errors 
like in the case of $\alpha$-p and t-p correlations) 
the result changes only marginally.
For a fixed source radius, 
the most probable time order of the emission is found as follows 
(cf. table\,\ref{time_sequence_p_d_a_t_3he}): 
On the average, protons are emitted first 
whereas $^3$He particles are emitted last (after about 11~fm/c). 
The other particles show up in between. Deuterons are emitted  
about 6~fm/c after protons. 
After that, the $\alpha$ particles are emitted and then
the tritons follow. However, the differences between the 
emission times of these species are found in the order of 1~fm/c 
only (note: 3~fm/c~=~10$^{-23}$~s).
\begin{table}[b]
\vspace*{10bp}
\caption{The 2nd row gives the emission times (relative to that of the 
protons) for fixed source radii $R_i=R_0$
of d, t, $^3$He, and $\alpha$ particles
as derived from a weighted regression of the ten linear equations 
for the pair-wise time delays given in 
table \,\protect\ref{pair_wise_time_delays}. The 3rd row gives the 
complementary information of different source radii 
(normalized to that of the proton source $R_{\mbox p}=R_0$)
if the emission takes place at the same time instant. 
\label{time_sequence_p_d_a_t_3he}
	}
{\setlength{\tabcolsep}{1.4mm}     
\begin{tabular}{ ccccc }
\hline
\protect\\
\vspace*{6bp}
   &  d & $\alpha$ & t & $^3$He  
\protect\\[1bp] \hline
\protect\\
\vspace*{6bp}
{\large 
$\frac{ \tau_i- \tau_{\small \mbox p}}{fm/c}$ } & $6.3\pm 0.8$ & $7.7\pm 0.9$ & $8.5\pm 0.9$ & $11.1\pm0.8$
\protect\\[1bp] \hline
\protect\\
\vspace*{6bp}
$R_i/R_{\mbox p}$ & $0.63\pm 0.04$ & $0.54\pm 0.05$ & $0.53\pm 0.05$ & $0.44\pm 0.04$  
\protect\\[1bp] \hline
\protect\\ 
\end{tabular}
}
\end{table}

As a consequence of the duality of space and time coordinates,  
the time delays can be transformed into position differences. 
When assuming a common emission time for all particles,  
the time sequence translates into the emission from different source radii   
$R_i$ according to Eq.\,(\ref{tau_star}) as a consequence of the radial flow. 
A very similar procedure as described above for the emission times 
at unique source size 
leads to the average radii for the emission at the same time instant. The results 
are summarized in the 3rd row of table\,\ref{time_sequence_p_d_a_t_3he}.
Now, the protons, 
which are emitted earliest in the time-ordered picture,   
would come from the most extended source ($R_{rms}=\sqrt{3} R_0 \simeq 10$~fm) 
whereas the clusters (with the exception of $\alpha$ particles, 
which obviously play a special role) are emitted from source  
radii which decrease with increasing mass. 
Thus, the size of the emission sources of the 
d, $\alpha$, t and $^3$He particles 
would be about 63\%, 54\%, 53\% and 44\% of that of the protons, respectively.

This finding is in qualitative agreement with the results of an  
earlier investigation \cite{Petrovici} of the cluster
formation process in central Au+Au collisions at 250 $A\cdot$MeV. There, the 
authors compare the experimental data with predictions of a model which   
describes the hydrodynamic isotropic expansion 
of an ideal nucleonic gas and the clustering by statistical disassembly. 
The model predicts a breakup which, with elapsing time, starts at the exterior 
and evolves to smaller radii. 
The heavier fragments are found to arise from smaller 
source radii than the light particles.

The $\alpha$ particles do not follow completely the systematic trend 
established above. One reason for this violation of the 
emission order with increasing mass might lie in the fact that this  
(strongly bound) particle species, at least partially,  
is made of nucleons which were initially correlated either in the 
target or projectile nucleus \cite{Gossiaux}. Such preformed clusters 
would carry to some extent a memory of the entrance channel.

We have studied, whether the emission time 
difference of protons and composite particles can be explained 
by calculations we have performed with the
IQMD transport model \cite{Bass}.
For a central ($b<4$~fm) reaction of Ru+Ru we have investigated the 
distribution of the time instants at which the particles 
pass through the surface of a sphere in coordinate space. 
Only particles coming from the participant zone 
have been selected by a cut on the c.m. polar angle 
$\vert \tan{\Theta_{cm}} \vert>1$.  
Since correlation functions are sensitive 
to particle pairs with small relative momenta only, 
in addition, we have to demand that the velocities of both  
particles be the same. The time distributions of all  
the light charged particles are found almost symmetric and exhibit     
- within the statistical errors - identical mean values. 
Alternatively, the cross check 
of the radial distributions of particles with equal velocities 
at a unique time reveals no differences of the mean radii. 
This finding is not surprising since most of the transport codes 
generate composite particles by coalescence.  
Indeed, such a method provides the composite particles with the same 
space-time distribution as the nucleons of equal velocity.

\section{Summary} \label{summary}
In conclusion, we have presented 
experimental correlation functions 
of nonidentical light charged particles produced in central 
collisions of Ru(Zr)+Ru(Zr) at 400 $A\cdot$MeV. 

For the first time an 
emission order of p, d, $\alpha$, t, and $^3$He  
particles has been set up by comparing correlations of 
particles with relative momenta parallel and anti-parallel to the 
center-of-mass velocity of the pair. Collective radial expansion of the  
participant zone leads to an apparent reduction of the source 
radius and to shifts of the emission times. 
Correcting for both effects typical time delays of a few fm/c 
were obtained. 
The deduced space-time differences of the light-charged-particle 
emission sources allow two complementary interpretations. 
If the source radius is fixed  
the composite particles are emitted at later times than protons. 
Alternatively, if the emission time is fixed, 
the clusters are emitted from smaller sources than protons.  
As a result of the 
duality of space and time coordinates, these two scenarios cannot be   
distinguished from each other.

\end{document}